\documentclass[iop]{emulateapj}
\newcommand{\HI}{\textsc{H}{\,{\textsc i}}}

\slugcomment{Accepted for publication in ApJ.}

\shorttitle{Dynamical Modeling with Identikit}
\shortauthors{Privon et al.}

\begin{document}

\title{Dynamical Modeling of Galaxy Mergers using Identikit}

\author{G. C. Privon\altaffilmark{1,}\altaffilmark{2}}
\author{J. E. Barnes\altaffilmark{3}}
\author{A. S. Evans\altaffilmark{1,}\altaffilmark{4}}
\author{J. E. Hibbard\altaffilmark{4}}
\author{M. S. Yun\altaffilmark{5}}
\author{J. M. Mazzarella\altaffilmark{6}}
\author{L. Armus\altaffilmark{7}}
\author{J. Surace\altaffilmark{7}}

\altaffiltext{1}{Department of Astronomy, University of Virginia, Charlottesville, VA 22904, \email{gcp8y@virginia.edu}}
\altaffiltext{2}{Visiting Graduate Student Research Fellow, NASA Infrared Processing and Analysis Center, California Institute of Technology, Pasadena, CA 91125}
\altaffiltext{3}{Institute for Astronomy, University of Hawaii, Manoa, HI}
\altaffiltext{4}{National Radio Astronomy Observatory, Charlottesville, VA, 22904}
\altaffiltext{5}{Astronomy Department, University of Massachusetts, Amherst, MA, 01003}
\altaffiltext{6}{NASA Infrared Processing and Analysis Center, California Institute of Technology, Pasadena, CA 91125}
\altaffiltext{7}{Spitzer Science Center, California Institute of Technology, Pasadena, CA 91125}

\begin{abstract}
We present dynamical models of four interacting systems: NGC 5257/8, The Mice, the Antennae, and NGC 2623. The parameter space of the encounters are constrained using the Identikit model-matching and visualization tool. Identikit utilizes hybrid N-body and test particle simulations to enable rapid exploration of the parameter space of galaxy mergers. The Identikit-derived matches of these systems are reproduced with self-consistent collisionless simulations which show very similar results. The models generally reproduce the observed morphology and \HI\ kinematics of the tidal tails in these systems with reasonable properties inferred for the progenitor galaxies. The models presented here are the first to appear in the literature for NGC 5257/8 and NGC 2623, and The Mice and the Antennae are compared with previously published models. Based on the assumed mass model and our derived initial conditions, the models indicate the four systems are currently being viewed 175-260 Myr after first passage and cover a wide range of merger stages. In some instances there are mismatches between the models and the data (e.g., in the length of a tail); these are likely due to our adoption of a single mass model for all galaxies. Despite the use of a single mass model, these results demonstrate the utility of Identikit in constraining the parameter space for galaxy mergers when applied to real data.
\end{abstract}


\section{Introduction}
\label{sec:Introduction}

There is increasingly strong evidence that galaxy mergers are an important aspect of galaxy evolution. This evidence includes the hierarchical formation of structure observed in $\Lambda$CDM cosmological simulations \citep[e.g.,][]{Springel2005,Boylan-Kolchin2009,Angulo2012}, as well as direct observations of high redshift systems with morphological irregularities consistent with merging or interacting systems \citep[e.g.,][]{Engel2010b,Bussmann2012}. As is observed at high-redshift, the most extreme star-forming systems and many of the most luminous active galactic nuclei (AGN) in the local Universe are associated with mergers \citep[e.g.,][]{Sanders1988,Bahcall1995}. Galaxy mergers can redistribute existing stars as well as trigger the formation of new stars through compression of gas clouds and the resuling increased gas density.

Numerical models for merging systems were first convincingly demonstrated by \citet[][hereafter TT72]{Toomre1972} and have been updated in subsequent decades. Historically, finding an accurate dynamical model has been a very time consuming process, owing to the need for iterative matching of self-consistent simulations. This difficulty has severely limited the number of systems with dynamical models. For a relatively recent summary of systems with dynamical models, see \citet[][table 2. Hereafter BH09]{Barnes2009}. 

Identikit (BH09) utilizes a novel approach to simplify matching of dynamical models to mergers. This method, described in detail in Section \ref{sec:Matching}, utilizes hybrid test particle \& N-body simulations and a data visualization technique which allow rapid exploration of the parameter space for galaxy mergers. This will allow a sample of galaxy mergers with dynamical models to be assembled. Such a sample could be paired with detailed simulations incorporating gas physics, star formation, and feedback to enable a comparison of the observed star formation and AGN properties in mergers with the characterization of those phenomena in simulations. This can test and lead to an improved understanding of the sub-grid physics necessary to most accurately incorporate this physical activity in numerical models.

At a minimum, a dynamical model\footnote{Here, we take ``dynamical model'' to include the minimum number of geometric and orbital parameters necessary to specify a galactic encounter. This includes the orbital parameters (mass ratio, orbital eccentricity, and pericentric separation), disk orientations relative to the orbital plane, and factors which relate to the observations (time of viewing, viewing angles, and scalings to physical units). With this definition, prescriptions for gas physics, star formation, AGN activity, etc. can be paired with a dynamical model to study/predict the appropriate phenomenon.} of a merger should reproduce the morphology and kinematics of the system. For the majority of systems modeled to date, this primarily means matching the tidal features formed during close passage of two galaxies with dynamically cold components. For prograde passages (disk spins aligned with the angular momentum of the orbit), quasi-resonant tidal interactions create tails \citep[TT72;][]{D'Onghia2010} such as those seen in the Antennae, The Mice, and other systems \citep{Arp1966,Schweizer1982,Hibbard2001,Hibbard1996}. These tails, once created, evolve ballistically and are relatively unaffected by the later evolution of the main galaxy bodies. Thus they provide the best constraints on the initial conditions of the encounter. 

An ensemble of dynamical models of mergers can enable a study of merger driven activity constrained in more detail than a statistical comparison of data and simulations. The difficulty of arriving at dynamical models for mergers has thus far precluded the assembly of a large sample of observationally-matched dynamical models. The aim of this paper is to demonstrate the feasibility of obtaining dynamical models with the Identikit technique.

The paper is organized as follows: we begin by describing the sample of mergers in Section \ref{sec:sample}. In Section \ref{sec:Matching} the Identikit hybrid and self-consistent simulations are described, and the results of matching the simulations to observations with Identikit are given in Section \ref{sec:Results}. We use a cosmology with $H_o=73$ km s$^{-1}$ Mpc$^{-1}$, $\Omega_{M}=0.27$, and $\Omega_{vac}=0.73$, unless otherwise noted.

\section{The Sample and Data}
\label{sec:sample}

The four galaxy mergers in this initial sample are NGC 5257/8, The Mice, Antennae, and NGC 2623. They were selected on the basis of having prominent tidal tails, indicative of prograde encounters (hence easier to match to simulations\footnote{In their study of simulated mergers BH09 found that simulated prograde and retrograde encounters were modeled with similar success rates. Also, preliminary work on expanding this sample to retrograde encounters shows observed retrograde systems can be successfully modeled using this technique (Privon et al. \emph{in prep}).} ) and also on the availability of high quality optical imaging and \HI\ observations. The detailed morphological and kinematic information from these data is critical in matching a dynamical model to each system using Identikit. The systems appear to span a range of merger phases, based on their optical morphology. Additionally they exhibit a range in infrared luminosity and therefore star formation rates \citep[e.g.,][]{Sanders2003,Kennicutt1998}. 

All four systems are well studied mergers. NGC 5257/8 and NGC 2623 are part of the Great Observatories All-sky LIRG Survey \citep[GOALS,][]{Armus2009}, which consists of imaging and spectroscopy on all LIRGs in the IRAS Revised Bright Galaxy Sample \citep[RBGS,][]{Sanders2003} from radio wavelengths through X-rays. Where appropriate we leverage this detailed dataset to assist in the development and interpretation of our models. The other two systems, The Mice and the Antennae, are also well studied systems with a wealth of existing data.

NGC 5257/8 and NGC 2623 do not have previously published dynamical models while The Mice and the Antennae have existing dynamical models (see Sections \ref{ssec:Mice} and \ref{ssec:Antennae}). The aim of this paper is not to determine the ``best'' of competing dynamical models, rather it is to demonstrate the feasibility of Identikit in obtaining valid models. 

For all sources, redshifts were retrieved from NED, D$_L$ was obtained from NED's ``cosmology corrected quantities'', and L$_{IR}$ was computed using D$_L$, the IRAS fluxes from NED's photometric database, and the prescription in \citet[][Table 1]{Sanders1996}. Some properties of these objects are summarized in Table \ref{table:sample}. A comparison of L$_{IR}$ determined by SED fitting versus estimates from IRAS fluxes by \citet{U2012} found both methods to provide similar results. 

\begin{deluxetable*}{llccc}[h!]
\tablecaption{Sample Description}
\tablehead{\colhead{Source} & \colhead{Other Names} & \colhead{$z$} & \colhead{D$_{L}$} & \colhead{L$_{IR}$} \\
\colhead{} & \colhead{} & \colhead{} & \colhead{\emph{Mpc}} & \colhead{\emph{$\times10^{11}$ L$_{\odot}$}}}
\startdata
NGC 5257/8 & Arp 240, VV 55, IRAS F13373+010 & 0.0225\phn\phn & 98.3 & 3.9\phn \\
The Mice & NGC 4676A/B, Arp 242, VV 224, IRAS 12437+3059 & 0.022049 & 96\phd\phn & 0.79 \\
Antennae & NGC 4038/9, Arp 244, VV 245, IRAS 11593-1835 & 0.005688 & 28.4 & 1.1\phn \\
NGC 2623 & Arp 243, VV 79, IRAS 08354+2555 & 0.018509 & 80.6 & 3.6\phn
\enddata
\label{table:sample}
\tablecomments{Col 1: Source name as used in this paper. Col 2: Other names used in the literature. Col 3: redshift obtained from NED. Col 4: Luminosity distance from NED's Cosmology-Corrected Quantities. Col 5: Infrared luminosity from IRAS observations. See text for details.}
\end{deluxetable*}

In the following subsections we describe the properties of the systems in our sample based on previous studies. The dynamical models presented here utilize the \HI\ observations and optical imaging. Other data sets and results are included as context for the merger-driven activity in each system.

\subsection{NGC 5257/8}
\label{ssec:NGC5257}

The pair NGC 5257 and NGC 5258 both show clear evidence for tidal distortion, with tidal tails and enhanced spiral features visible in stellar light. Neither nucleus shows evidence for the presence of a strong AGN. Photometric estimates and SED modeling of the system provide stellar mass estimates of $1-3\times10^{11}$ M$_{\odot}$ \citep{Howell2010,U2012}.

HST ACS imaging of NGC 5257/8 was obtained as part of the GOALS observing program (pid10592 PI: Evans; Evans et al. in prep) in the F435W and F814W filters. Over 800 star clusters are visible in the image (Vavilkin et al. in prep). The majority of these clusters are in the main body and spiral arms, but there are also clusters in the more extended tidal features. Using GALEX UV observations in combination with the FIR luminosity, \citet{Howell2010} find total SFRs of $35.6$ and $36.0$ M$_{\odot}$ yr$^{-1}$ for NGC 5257 and 5258, respectively. 

Spitzer imaging of the system shows the dominant source of infrared emission to be off-nuclear, in sharp contrast to the general behavior of LIRGs where activity is frequently concentrated in the nucleus \citep{Haan2011}. These star forming regions lie coincident with the spiral arms which are likely enhanced by the tidal interaction. This may be leading to enhanced compression of the gas, beyond that typically occurring in spiral arms. Independent of the dynamical modeling, this system affords an opportunity to study extreme star formation in a region of the galaxy where influence from an AGN can be reliably ruled out.

\citet{Iono2005} presented VLA \HI\ observations, mapping the distribution of the atomic gas in the main bodies and the tails. The system has a total \HI\ mass of M$_{\HI}=3.3\times10^{10}$ M$_{\odot}$. The tidal tail in NGC 5257 is relatively short in projection, suggesting a weak tidal coupling (i.e., not a strongly prograde encounter) or a tidal disturbance predominantly along the line of sight. The weak \HI\ emission here may be due to a geometry for the system in which the \HI\ in the tail has been ionized by star formation in the disk of the galaxy (e.g., the base of the northern tail in the Antennae: \citet{Hibbard2001}). The prominence of the tidal tail in NGC 5258 is suggestive of a prograde interaction. The eastern tail has a greater extent in \HI\ than in starlight. This suggests the \HI\ in the tail was initially less tightly bound, i.e., in an extended disk of the type often seen in non-interacting disk galaxies.

The system has no previously published dynamical models. For input constraints to Identikit, we use the \HI\ observations from \citet{Iono2005} and the HST/ACS F814W images from Evans et al. (in prep). The \HI\ observations have an angular resolution of $\sim20\arcsec$ ($9.1$ kpc, FWHM) and a velocity resolution of $11$ km s$^{-1}$. The HST image has a resolution of $0.05\arcsec$. 

\subsection{The Mice}
\label{ssec:Mice}

The Mice is a well known interacting galaxy pair. The system received its name from the prominent tails which result in the system resembling two ``playing mice'' \citep{Vorontsov-Vel'Iaminov1958}. These prominent tidal tails suggest a prograde interaction. The straightness of the northern tail is consistent with the spin vector of the northern galaxy being near the plane of the sky. The relative length of the tidal tails suggests the galaxies are of roughly equal mass.

HST/NICMOS imaging of the component galaxies shows both have compact nuclei \citep{Rossa2007}. A comparison of NGC 4676A and NGC 4676B with HST/WFPC2 imaging is consistent with NGC 4676A being of a slightly later Hubble type than NGC 4676B \citep{Laine2003}.

Using Keck's multi-object spectrograph, \citet{Chien2007} observed 12 optically visible star clusters in the Mice, selected from HST/ACS imaging of the system \citep{deGrijs2003}. The observed clusters were predominantly young ($<10$ Myr), but 2 of the clusters had ages $\sim170$ Myr. 

This merger system has been the subject of many dynamical studies \citep[e.g., TT72;][]{Mihos1993,Gilbert1994,Sotnikova1998,Barnes2004}. The existing models have orbital parameters and disk inclinations which are in agreement with each other.

For our model matching we utilize \HI\ data and a deep optical R-band image from \citet{Hibbard1996}. The \HI\ data has a resolution of $20\arcsec$ ($8.9$ kpc) and a velocity resolution of $43.1$ km s$^{-1}$. The total \HI\ mass in the system is $7.5\times10^9$ M$_{\odot}$. The optical image was taken in $1.9\arcsec$ seeing; this resolution is sufficient for our analysis.

\subsection{Antennae}
\label{ssec:Antennae}

The Antennae consists of two distinct galaxies (NGC 4038 and NGC 4039), each with long tidal tails which cross close to their respective disks. The ``overlap region'' of the disks contains most of the molecular gas \citep{Wilson2000} and is the primary source of the far-infrared luminosity \citep{Mirabel1998,Xu2000}, though it is not clear if the activity in this region is due to physical contact between the two disks or if they are seen as overlapping in projection. Neither nucleus has evidence for a strong AGN. As with the previous systems, the tidal tails suggest prograde interactions for both disks. 

The Antennae has been the subject of many papers which analyze its dynamical properties \citep[e.g., TT72,][]{vanderHulst1979,MahoneyJ.M.;BurkeB.F.;vanderHulst1987,Barnes1988,Mihos1993,Karl2010}. These various models generally reproduce the morphology and/or the kinematics of the large scale features. The more detailed hydrodynamic model of \citet{Karl2010} appears to match the smaller scale features as well as providing a natural explanation for the ``overlap region'' as a region of physical contact between the disks. However, this model occupies a different region of parameter space when compared to previous efforts. We discuss the differences between our derived model and previously published models in Sections \ref{sssec:MiceModels} and \ref{sssec:AntennaeModels}.

\HI\ and an optical R-band image presented in \citet{Hibbard2001} were used to constrain the dynamical model. The \HI\ observations were taken with the VLA and cover the system at an angular resolution of approximately $10\arcsec$ ($1.3$ kpc) and a velocity resolution of $5.21$ km s$^{-1}$. The optical image has $1\arcsec$ seeing.

\subsection{NGC 2623}

NGC 2623 is, by visual inspection, a late stage merger. HST/ACS images of the system show a single nucleus and prominent $20-25$ kpc tidal tails extending to the East and West \citep{Evans2008}. The stellar mass of the system is estimated at $6\times10^{10}$ M$_{\odot}$ and $\sim2\times10^{10}$ M$_{\odot}$ based on photometric and SED estimates, respectively \citep{Howell2010,U2012}.

The merger remnant has $\sim100$ unresolved star clusters \citep{Evans2008}. Assuming negligible to moderate extinction, the HST/ACS two filter broadband photometry of the clusters is consistent with ages of $1-100$ Myr. \citet{Rothberg2007} give an alternate estimate for the age range of $100-250$ Myr, using an additional filter in the analysis.  Detailed spectroscopy of 18 clusters by \citet{Chien2009} found cluster ages between $3-250$ Myr, with most of the observed clusters younger than $100$ Myr. The overall star formation rate in the system is $\sim70$ M$_{\odot}$ yr$^{-1}$ \citep{Howell2010}.

The merger core hosts an AGN as determined from the hard X-ray spectrum \citep{Maiolino2003,Evans2008} and the presence of the [Ne V] $14.3~\mu$m line \citep{Evans2008,Petric2011}. However the energetics of the system are dominated by star formation. \citet{Lonsdale1993} detected a high brightness temperature (T$>10^7$ K) radio core consistent with an AGN, but contributing less than $10\%$ of the total radio flux. This suggests the bulk of the radio flux is associated with star formation.

We utilize \HI\ observations obtained with the VLA in both the D and C array configurations \citep{Hibbard1996a}, providing an angular resolution of $16\arcsec$ (6 kpc). The \HI\ shows a double tail morphology, with somewhat more emission from the western tail. The line of sight velocity of the Eastern tail increases monotonically from the middle of the tail to the end. In contrast, the Western tail velocity initially decreases with increasing radius. This tail then loops around in both position and velocity, curving towards the North and back around East, as well as reversing the line of sight velocity.

We have measured the flux of the \HI\ emission as: $0.71$ Jy km s$^{-1}$. At D$_L=80.6$ Mpc and taking into account the fact that we are missing flux due to \HI\ absorption against the radio continuum, this corresponds to M$_{\HI}>2.3\times10^{8}$ M$_{\odot}$. 

The \HI\ absorption is seen against the radio continuum which likely originates from the strong star formation in the system. At this resolution the absorption is unresolved, and the radio continuum may be smaller than the $16\arcsec$ beam. The absorption occurs almost the entire $450$ km s$^{-1}$ bandwidth of the VLA observations, which agrees with single-dish \HI\ observations with the GBT (Frayer et al. \emph{in prep}). Taking the NVSS flux of $95.7$ mJy \citep{Condon1998} we estimate the total absorbing column: N$_{\rm H}=3.6\times10^{21}$ cm$^{-2}$ (for a covering factor of 1 and a spin temperature T$_{\rm spin}=100$ K). 

Optically, the main body of the galaxy appears to have only a single nucleus \citep{Evans2008,Haan2011}, implying that higher resolution is required to resolve the nuclei of the progenitor galaxies, or the nuclei have already coalesced. In this case, one might expect the most tightly bound portions of the tails (at the smallest radii) to have reached apocenter and begun falling back into the system. Unfortunately, due to the strong absorption in the center, we are unable to use \HI\ to probe for the existence of a velocity reversal at small distances which would suggest infall. Deep optical spectra of the tails to determine the kinematic properties using stellar absorption lines would be an interesting confirmation of this.

To complement the \HI\ data with constraints on the stellar distribution, we utilize the HST/ACS F814W image from \citet{Evans2008}.

\section{Matching of Dynamical Models}
\label{sec:Matching}

Specifying a model of a disk-disk galaxy merger requires a minimum of 16 parameters. These include orbital parameters, galaxy disk inclinations, the viewing direction, and scalings to physical units. The orbital parameters are: mass ratio of the two galaxies ($\mu$), pericentric separation of the idealized Keplerian orbit of point masses\footnote{The model galaxies are initially set on a Keplerian orbit, but will deviate from this orbit shortly before first pericenter passage.} ($p$), and the eccentricity of the orbit ($e$).  The orientation of each disk is specified by two angles, the angle between the normal of the orbital plane and the unit vector of the disk's angular momentum (inclination, $i$), and the longitudinal orientation of the angular momentum unit vector\footnote{We define $\omega$ for the idealized Keplerian orbit, where the angle is the longitudinal argument relative to the line of nodes at pericenter. Due to the more rapid orbital decay of the extended mass distribution, pericenter will occur earlier than in the idealized case. This will cause the true $\omega$ value to differ slightly from the Keplerian approximation.} (argument of periapse, $\omega$). For an illustration of these angles, see TT72 Figure 6a.

The above parameters are sufficient to describe the dynamical evolution of the system. To specify an observation, additional parameters must be given, including: three angles corresponding to the viewing angle of the merger ($\theta_X,~\theta_Y,~\theta_Z$), a length scaling ($\mathcal{L}$), a velocity scaling ($\mathcal{V}$), and the time of viewing\footnote{We set the idealized Keplerian orbit to have first pericenter passage at $t=2$ time units.} ($t$). $\mathcal{L}$ and $\mathcal{V}$ are then used to determine the mass scaling factor $\mathcal{M}$ for the system: $\mathcal{M}\propto \mathcal{V}^2L$. These scaling factors can also be used to arrive at a time scaling factor to convert the simulation time into physical units ($\mathcal{T}\propto\mathcal{L}/\mathcal{V}$). The final three parameters are the location of the system's center of mass ($X$, $Y$, $z$).

Identikit \citep[BH09;][]{Barnes2011a} combines the use of hybrid self-consistent halos and test particle disk simulations (described in Subsection \ref{ssec:test}) with a visualization technique (described in Subsection \ref{ssec:datavis}) to facilitate rapid comparison of seven dimensional models (x, y, z, v$_x$, v$_y$, v$_z$, t) with three dimensional data ($\alpha$, $\delta$, v$_z$). Here we limit ourselves to the ``Identikit 1'' technique of BH09, which relies on manual investigation of parameter space. 

\subsection{Test Particle Simulations}
\label{ssec:test}

Identikit simulations are created by first selecting values for $\mu$, $r_{peri}$, and $e$. Then self-consistent galaxy models consisting of a spherical dark matter halo \citep[NFW profile,][]{Navarro1996}, spherical bulge \citep[Hernquist profile,][]{Hernquist1990}, and a ``spherical'' disk are generated using the Zeno software package\footnote{\url{http://www.ifa.hawaii.edu/faculty/barnes/zeno/} and \url{http://www.ifa.hawaii.edu/faculty/barnes/software.html}}. These ``live'' N-body galaxies are then populated with spherical swarms of massless test particles on circular orbits, representing all possible disk orientations. For details of the construction of these simulations see BH09. 

These hybrid N-body and test particle models are placed on an orbit corresponding to $p$ and $e$, and then evolved using standard numerical techniques \citep[a treecode, e.g.,][]{Barnes1986}. The positions and velocities of the test particles, which respond to the changing potential defined by the massive components, are saved at many snapshots throughout the observations. After the fact, specific disk orientations ($i$,$\omega$) can be extracted by selecting only those test particles with the appropriate initial angular momentum vectors.

By running simulations with different values of ($\mu$, $p$, $e$) a large grid of models can be stored. Test particle disks from these simulations can then be loaded on demand from within the Identikit interface and compared to data.

In general, matching simulations to data is made easier by the existence of tidal tails, which evolve ballistically once created and therefore retain memory of the initial encounter passage. Their global morphology and kinematics are not strongly modified by self-gravity and so should be well-matched by test-particles. In contrast, strong spiral arms and other self-gravitating features will not be well reproduced by the test particles.

\subsection{Data Visualization}
\label{ssec:datavis}

A key aspect of Identikit is the interface to display position and velocity information derived from data cubes simultaneously with results from simulations. The standard view is a 4 panel display (e.g., Figure 1 of BH09, Figure \ref{fig:idkN5257} of this paper), showing the sky view ($\alpha-\delta$, top left), two position-velocity views ($v$-$\alpha$ \& $\delta$-$v$, bottom left and top right), and a ``top down'' view ($\alpha$-$z$, bottom right).

The subset of test particles from a specific instance ($t$, $\mu$, $e$, $p$ $i$'s, $\omega$'s) of a hybrid simulation is loaded from the grid of models described in Section \ref{ssec:test} and overlaid on the sky-plane and position-velocity diagrams. Identikit can move through a grid of pre-computed models, stepping through time, pericentric separation and mass ratio. Disk orientations can be changed with the display updating in real-time. Scalings and the viewing direction can also be changed in real-time to match the data.

The data constraints are provided via three of the four panels described above. The source of these constraints may vary according to the available data. We generally employ sky-plane visualizations using deep optical images overlaid with contours of the \HI. The \HI\ and stellar tails may not be co-spatial for reasons including: different scale lengths of the stellar and gaseous disks in the progenitors, ionization effects \citep[e.g.,][]{Hibbard2001}, or dissipation from an extended gas disk \citep{Mihos2001}. Using both \HI\ and starlight ensures the effect of the tidal interaction is accurately visualized. We utilize red optical filters in order to avoid being biased by young stellar populations. Extinction is generally low in the tails so the extended morphology is relatively unbiased by the use of optical images. The resolution of the \HI\ is the limiting factor in terms of kinematic constraints but is sufficient for model matching. The higher resolution optical images are not degraded to the resolution of the \HI, as they may provide additional posterior constraints on the match (e.g., the width of the tidal tails, the match of self-gravitating features such as spiral arms).

The position-velocity diagrams were created using projections of the \HI\ data cube, taken along either the $\alpha$ or the $\delta$ axes. An image with the maximum voxel value along the specified axis was used. This method tends to provide a more complete view of the spatial distribution of the emission, as faint emission isn't averaged together with noise. A consequence of this is that the projections don't represent the total emission at that point in $\alpha-\delta-v$ space. Therefore, comparing the data value in a given location of a projection with the density of simulation particles in that same region is not a reliable metric for evaluating the accuracy of a match.

\subsection{Identikit Matching}

Before matching in Identikit, the available parameter space was constrained as much as possible using a priori information. The general disk orientations (prograde vs. retrograde) can be guessed at using the length and prominence of tidal features. For the orbit, $e\sim1$ is assumed based on cosmological considerations: the galaxies should have a high-eccentricity orbit with a period comparable to the Hubble Time. As noted above, the mass ratio $\mu$ can be estimated from the relative luminosities of each galaxy, if they are still separable, or from the relative length of the tidal tails. For this work, we limited ourselves to $\mu=1$ or $\mu=2$ mass ratios.

With rough parameter estimates for each source, the data projections were loaded into Identikit and test particles with a specific initial angular momentum overplotted (with that initial angular momentum corresponding to a requested initial disk orientation). Disk orientations, time since pericenter, viewing angles and length/velocity scalings were varied until an acceptable match to the morphological and velocity information were obtained. Typically, approximate disk orientations were established and then the viewing angles explored until a near match of the tidal features is obtained. The match is then refined by alternately improving the disk angles and the viewing angles. Finally, length and velocity scalings were adjusted to obtain a suitable match to the velocity and spatial extents. These matches were determined by eye and were judged on the basis of having test particles in regions of the data projections where there is either stellar emission or \HI\ emission. The presence of test particles in regions without emission is not always an issue as emission could be missing due to effects not included in our simulations (e.g., ionization of the gas).

Despite the lack of a dissipational component in the hybrid simulations, matching these test particles to the kinematics of gas is still valid in the tails. This is because gas in the tails does not experience significant dissipation as it moves outwards.

A mass scaling ($\mathcal{M}$) can be derived from the velocity and length scalings, which allows one to estimate the dynamical mass of the system\footnote{Estimates of the baryonic mass require simulations with a self-consistent treatment of the disk (e.g. Section \ref{sec:selfcon}).}. The dynamical masses estimated from these simulations should be considered a lower limit to the true dynamical mass of a system. Lower values of the dynamical mass would be inconsistent with the velocity width and sizes observed in the objects. Due to the logarithmic dependence of the gravitational potential on the halo mass, increasing the halo mass would not significantly alter the velocity and length scalings. In other words, once the minimum mass threshold is reached, the sensitivity of these scalings to the dynamical mass is reduced. 

For evaluation of test particle matches, the most weight is given to the extended tidal features. As these evolve ballistically after the initial passage, self-gravity should be relatively unimportant and will be suitably matched by test particles. Features such as spiral arms, which require self-gravity, will not be reproduced by these massless test particles. As a result, mis-matches in the main bodies of the galaxy between the observations and the hybrid N-body and test particle simulations are to expected. Indeed, taking the parameters from an Identikit match and re-simulating them with a self-gravitating disk is a useful discriminant between models, particularly in systems lacking prominent tidal features.

BH09 tested Identikit by matching 36 simulated mergers without prior knowledge of a subset of the parameters (the mass model, mass ratio, and orbital eccentricity were fixed, all other parameters were randomly chosen as described in the paper). The viewing directions and spin orientations were typically recovered to within $12^{\circ}$ and $18^{\circ}$, respectively. The time since pericenter was also well-recovered, with no obvious biases in the determination of that parameter. The determination of the pericentric separation, while not showing an offset, did show the largest scatter. The determination of the pericentric separation is likely uncertain by a factor of two. The only parameter which shows a noticeable bias in the best fit values compared to the true values is the velocity scaling. However, this is easily explained by the lack of velocity dispersion in the test-particle method. The test-particle disks, which initially have zero velocity dispersion, are being used to match tails formed from disks with a non-zero velocity dispersion. The velocity scaling determined by test-particle matching would thus be inflated to compensate for this.

\subsection{Verification With Self-Consistent Simulations}
\label{sec:selfcon}

After matching a dynamical model using Identikit, self-consistent simulations were carried out to verify the solution, refine the model parameters, and ensure self-gravitating features are matched. All simulations were carried out using the Zeno software.

Model galaxies were created with a NFW dark matter halo and spherical bulge. In contrast with the Identikit method, a disk of gravitating particles was inserted with the specified ($i$, $\omega$) values. Our N-body simulations used 64k particles per galaxy: 8k bulge particles, 24k disk particles, and 32k dark matter halo particles. Eighty per-cent of the mass is in a dark matter halo and the remaining twenty per-cent is baryons (the bulge and the disk).  These galaxy models were placed on initially Keplerian orbits with $e$ and $p$ values taken from the Identikit solution, then evolved using a treecode. All galaxies use the same mass model: a scale length of $\alpha^{-1}=1/12$ for the exponential disk, bulge scale length of $a_b=0.02$, and halo scale length of $a_h=0.25$. The circular velocity at 3 disk scale lengths is approximately 1.26. These lengths and velocities can be converted to physical units using the scalings derived for each model. For the details of the construction of our N-body realizations see Appendix B of BH09.

Once the system has been evolved numerically, particles in the self-gravitating disks were loaded into the Identikit visualization system and compared with the same data projections used to constrain the test particle model. Minor adjustments were sometimes necessary, typically with the pericentric separation of the first passage. A model was considered sufficiently accurate if the self-consistent simulation reproduced the morphology and kinematics of the extended tidal features. 

Based on the comparison of Identikit matches to simulated mergers by BH09, we are confident that a visually good Identikit match should also provide angles which are reasonably close to the true values. In practice, the disk angles derived with test particles accurately reproduce the tidal features in self-consistent realizations of the merger, with minor ($<10^{\circ}$) adjustments sometimes necessary.

The baryonic mass of the galaxy can be estimated from the inclusion of the self-gravitating disk. The disks in our model galaxies are marginally unstable to bar formation and therefore are near the upper envelope of possible disk masses. Correspondingly, the baryonic masses given in the results are near the maximum plausible value for these mass models. The dynamical mass is a lower-limit, as with the test-particle simulations. The integrated mass-to-light ratio of the system is therefore a lower limit.

The models presented here are intended to reproduce the morphology and kinematics of the large scale structure, which they appear to successfully achieve. While these systems include gas, we have not included any dissipational component in the simulations. Such a component is necessary for studying the hydrodynamics of the system but its omission here should not affect the global fit, as the large scale features and orbital decay are not strongly affected by the inclusion of gas \citep{Barnes1996}.

\section{Results and Discussion}
\label{sec:Results}

\begin{figure}
\includegraphics[width=0.5\textwidth]{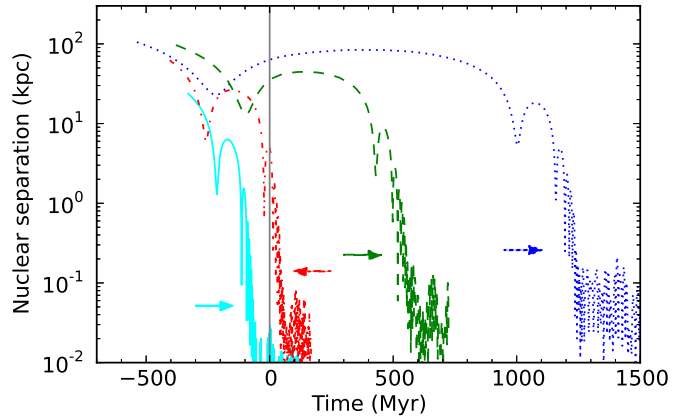}
\caption{True nuclear separation as a function of time for NGC 5257/8 (dotted blue line), The Mice (dashed green), Antennae (dash-dot red), and NGC 2623 (solid cyan). Time of zero is the current viewing time (solid gray vertical line). The time since first passages for these systems is $175-260$ Myr (cf. Table \ref{table:models}). Colored arrows mark the smoothing length in kpc for the corresponding system; this is effectively the spatial resolution of our simulations and the behavior of the curves on length scales smaller than the smoothing length is not reliable.}
\label{fig:decay}
\end{figure}

Dynamical models were successfully matched to the four systems selected. In all systems the prominent tidal features are traced back to prograde encounters, as was expected. Parameters for dynamical matches and scalings to physical units determined by Identikit and refined by subsequent self-consistent simulations are given in Table \ref{table:models}. All four systems are found to be consistent with equal-mass encounters ($\mu=1$). The orbital decay of the systems are shown in Figure \ref{fig:decay}. The modeling results for individual objects are discussed in the following subsections.

\begin{deluxetable*}{lcccccccccccc}[h!]
\tablecaption{Dynamical models derived from Identikit matching}
\tablehead{\colhead{System} & \colhead{$e$} & \colhead{$p$} & \colhead{$\mu$} & \colhead{($i_1$, $\omega_1$)} & \colhead{($i_2$, $\omega_2$)} & \colhead{$t$} & \colhead{($\theta_X$, $\theta_Y$, $\theta_Z$)} & \colhead{$\mathcal{L}$} & \colhead{$\mathcal{V}$} & \colhead{M$_{dyn}$} & \colhead{$t_{now}$} & \colhead{$\Delta t_{merge}$}\\
\colhead{} & \colhead{} & \colhead{} & \colhead{} & \colhead{} & \colhead{} & \colhead{} & \colhead{} & \colhead{(kpc)} & \colhead{(km s$^{-1}$)} & \colhead{($\times 10^{11}$ M$_{\odot}$)} & \colhead{(Myr)} & \colhead{(Myr)}}
\startdata
NGC 5257/8 & 1 & 0.625	 & 1 & (85$^{\circ}$, \phd65$^{\circ}$) & (15$^{\circ}$, 340$^{\circ}$)	& 3.38 & (126$^{\circ}$, \phd\phd-3$^{\circ}$, \phd\phd63$^{\circ}$) & 34\phn\phd & 204 & 9\phn\phd & 230 & 1200 \\
The Mice   & 1 & 0.375	 & 1 & (15$^{\circ}$, 325$^{\circ}$)& (25$^{\circ}$, 200$^{\circ}$)	& 2.75 & (\phd78$^{\circ}$, -44$^{\circ}$, -130$^{\circ}$) 	     & 39.5	  & 165 & 6.6	    & 175 & \phd775 \\
Antennae   & 1 & 0.25\phd& 1 & (65$^{\circ}$, 345$^{\circ}$)& (70$^{\circ}$, \phd\phd95$^{\circ}$)	& 5.62 & (-20$^{\circ}$, 283$^{\circ}$, \phd\phd-5$^{\circ}$)	     & 19.7	  & 265 & 8\phn\phd & 260 & \phd\phd70 \\
NGC 2623   & 1 & 0.125	 & 1 & (30$^{\circ}$, 330$^{\circ}$)& (25$^{\circ}$, 110$^{\circ}$)	& 5.88 & (-30$^{\circ}$, \phd15$^{\circ}$, \phd-50$^{\circ}$) 	     & \phd6.9	  & 123 & 0.6	    & 220 & \phd-80
\enddata
\label{table:models}
\tablecomments{$e$ -- orbital eccentricity, $p$ -- pericentric separation (simulation units), $\mu$ -- mass ratio, ($i_1$, $\omega_1$) ($i_2$, $\omega_2$) -- disk orientations (see text for description), $t$ - time of best match (simulation units, see text for description), ($\theta_X$, $\theta_Y$, $\theta_Z$) -- viewing angle relative to the orbit plane, $\mathcal{L}$ -- length scaling factor, $\mathcal{V}$ -- velocity scaling factor, M$_{dyn}$ -- estimate of the dynamical mass, $t_{now}$ -- time since first pericenter passage, $\Delta t_{merge}$ -- time until coalescence based on the assumed mass model.}
\end{deluxetable*} 

Assigning uncertainties to the model parameters in Table \ref{table:models} is somewhat subjective as the model fits are essentially qualitative and done by eye. We note there may be systematic uncertainties due to mismatches between our mass models and the true construction of these galaxies. Additionally, these parameters are correlated, often in nonintuitive ways. For example, selecting a later viewing time is generally compensated for with a larger pericentric separation and a change in scale factors. This will also require a modification of the viewing direction to produce the same kinematic and morphological signature. Despite these difficulties, there typically appears to be a single region of parameter space which provides a unique match. Uncertainities on the parameter values were quantified by modifying the parameters from the values in Table \ref{table:models}. Values within the range of uncertainties quoted here did not significantly degrade the fit. The vector directions of the disk spins should be accurate to within $15^{\circ}$; further deviations result in significant morphological and kinematic discrepancies with observations. Similarly the viewing direction appears to have a range of $20^{\circ}$ while still providing acceptable results. As noted above, the pericentric separation is anti-correlated with the viewing time; a factor of 2 uncertainty in $p$ is reasonable. Once the scaling factors are included to express the time in physical units, the time since first pericenter remains relatively stable, with acceptable matches lying within a $\pm15\%$ range. The mass ratio $\mu$ was fixed at $1$, but we estimate it could vary by up to $50\%$ for the systems modeled here.

Visualizations of the evolution of these dynamical models are available with the digital article\footnote{URL to be provided by ApJ} and will also be made available through NED via object name searches.

\subsection{NGC 5257/8}

\begin{figure*}
\includegraphics[width=\textwidth]{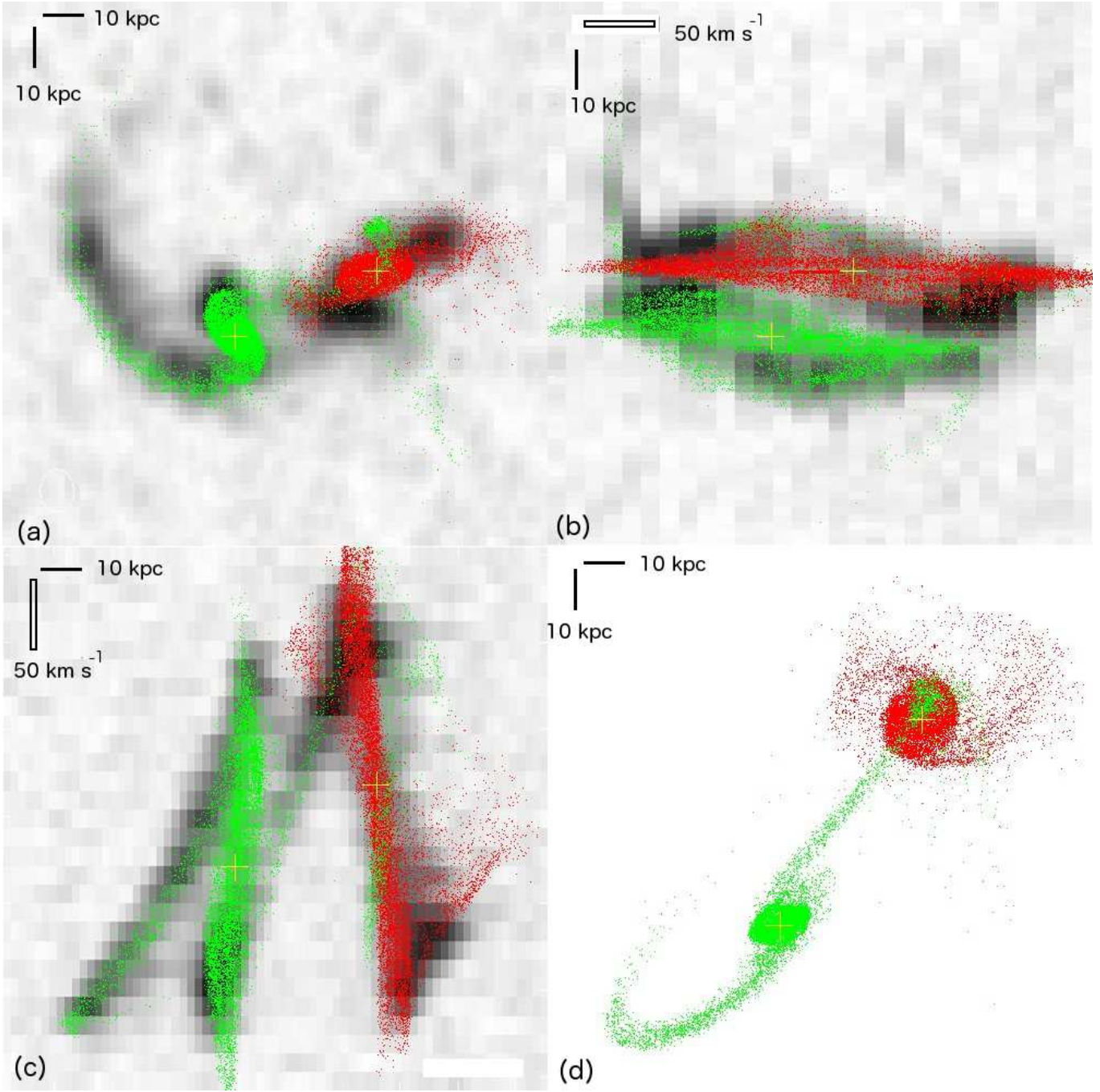}
\caption{Identikit visualization of a self-consistent model for NGC 5257/8, matched to the system. (a): sky view of the system ($\alpha$-$\delta$), (b) PV diagram ($v$-$\delta$), (c): PV diagram ($\alpha$-$v$), and (d) ``top-down'' view ($\alpha$,$z$). The sky view covers 133 kpc on a side and the velocity range is 650 km s$^{-1}$. In the relevant panels, the solid bar is 10 kpc and the box is $50$ km s$^{-1}$.  The \HI\ data is shown in grayscale, with the darker pixels corresponding to higher peak values along a vector through the data cube. Red and green points show collisionless baryonic particles from N-body realizations representing the galaxies NGC 5257 and NGC 5258, respectively. The yellow crosses represent the nuclei of each N-body realization.}
\label{fig:idkN5257}
\end{figure*}

\begin{figure}
\includegraphics[width=0.5\textwidth]{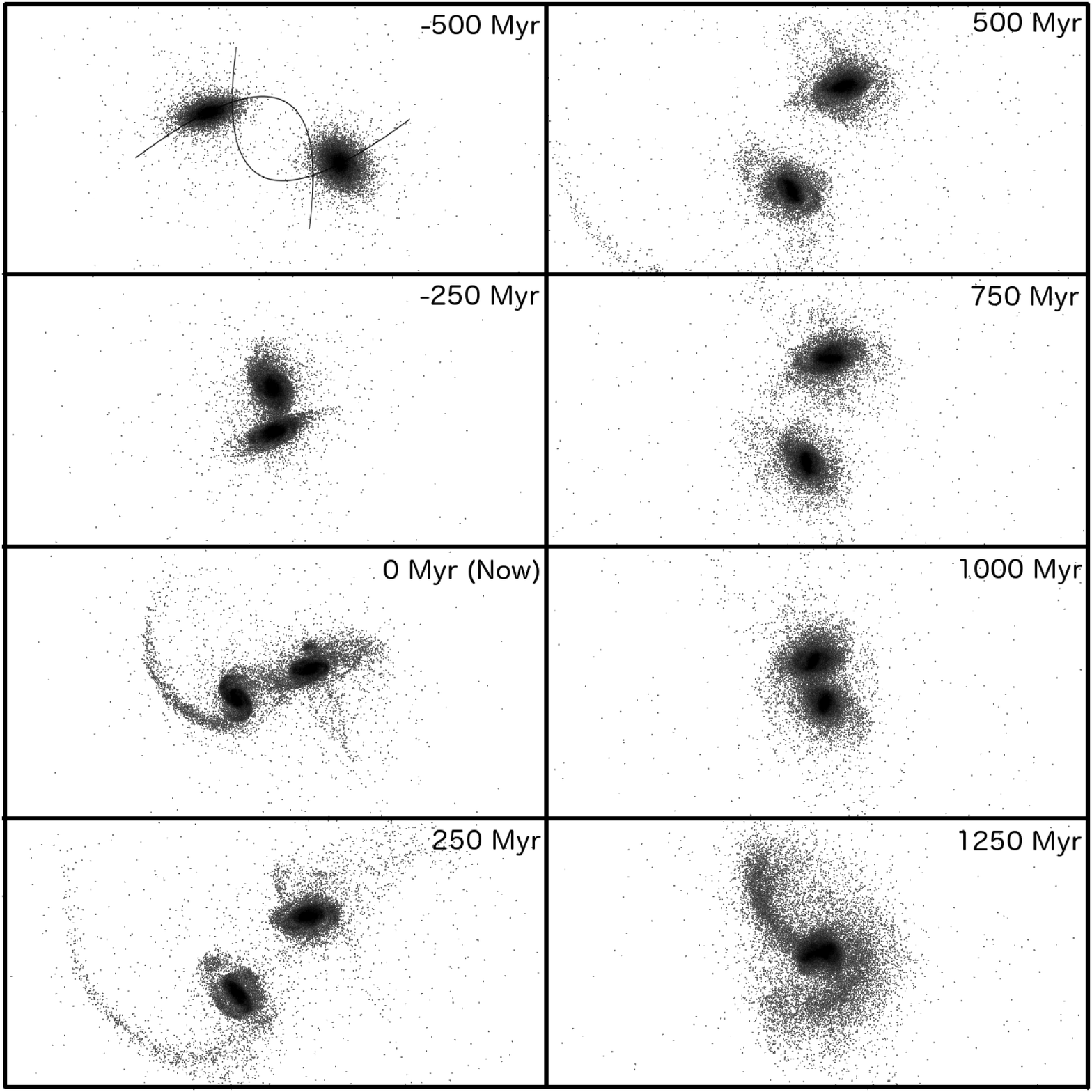}
\caption{Snapshots of the evolution of NGC5257/8 as a function of time based on the dynamical model presented here and shown in the sky plane. The stellar distribution is shown in grayscale in 375 kpc by 240 kpc boxes. The lines in the upper-left panel show the initial Keplerian orbit projected onto the sky plane. The times shown are relative to now (cf. Figure \ref{fig:decay}).}
\label{fig:N5257evol}
\end{figure}

As would be inferred from visual inspection, NGC 5257/8 is an early stage merger between two massive spiral galaxies (Figures \ref{fig:idkN5257} and \ref{fig:N5257evol}). The disk spin axis of NGC 5257 is nearly perpendicular to the orbital plane while the disk spin axis of NGC 5258 is only moderately inclined. The best-fit disk inclinations are ($85^{\circ}$, $65^{\circ}$). The system is viewed t$_{now}\sim220$ Myr after first pericentric passage. The galaxies are approaching apocenter but have not yet reached the point of greatest separation.

The model scaling factors suggest that the total system has baryonic mass $\sim1.8\times10^{11}$ M$_{\odot}$ and dynamical mass of M$_{dyn}\sim9\times10^{11}$ M$_{\odot}$. The baryonic mass determined here is consistent with estimates from photometry and SED fitting. The disks have scale lengths of $2.8$ kpc and a circular velocity of $257$ km s$^{-1}$ at 3 disk scale lengths\footnote{This circular velocity is the value if the disk were entirely rotation supported. These disks are partially dispersion supported, so the actual circular velocity is slightly lower.}. The idealized Keplerian orbit is a somewhat wide passage of $p\mathcal{L}=r_{peri}=21$ kpc (Figure \ref{fig:decay}). 

\begin{figure}
\includegraphics[width=0.5\textwidth]{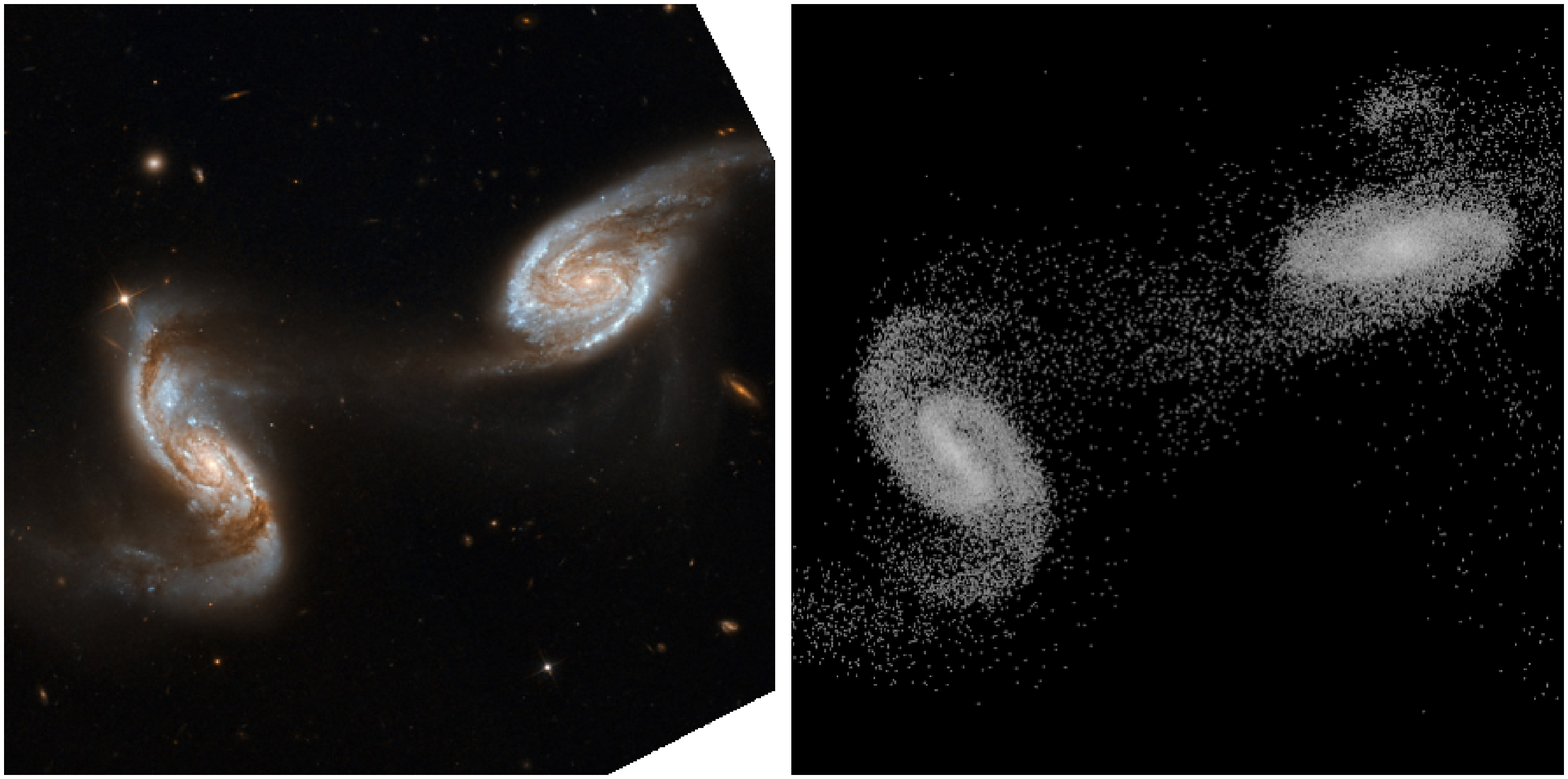}
\caption{Left: HST/ACS color image of NGC 5257/8 from Evans et al. (\emph{in prep}). Right: N-body realization of the stellar distribution from the dynamical model presented here. Note the general correspondence between the spiral features in the simulation and the observations.}
\label{fig:N5257comp}
\end{figure}

This dynamical model reproduces the kinematics of the tidal features, particularly the eastern tidal tail. Simulating this model with a self-gravitating disk shows that this encounter geometry is additionally able to reproduce the spiral structure seen in NGC 5258, and to a lesser degree of accuracy, the spiral structure in NGC 5257 (Figure \ref{fig:N5257comp}). These spiral features are generally not evident in the hybrid simulations, so they provide an extra test of a model's validity.

Roughly 30 clusters have been age dated in NGC 5257/8 using optical and UV colors obtained from HST imaging. The clusters can be broken into two populations: older, unembedded clusters whose ages are less than or equal to $240$ Myr and a group of clusters with uncertain ages due to unknown reddening (Manning et al. in prep). The upper envelope of ages is similar to the first pericenter passage, suggesting these clusters were formed while the gas experienced its strongest response to tidal forces. The second population of clusters has uncertain ages, but for modest amounts of extinction (A$_V\leq2$) these clusters would have ages of a few Myr. These are likely associated with ongoing gas compression. The clusters are predominantly located along the spiral arms in both galaxies, which have been enhanced by the tidal interaction. The close agreement of the cluster ages with the first passage and the distribution of clusters may be an interesting constraint on star formation prescriptions.

The dynamical model presented here has some minor deficiencies. For one, the short tail associated with NGC 5257 shows a greater extent in the model than is observed. This could be remedied with a smaller disk scale length. The tail associated with NGC 5258 is longer in \HI\ than in starlight, suggesting the gas and stellar disks have different scale lengths. A more accurate mass model may also reduce the ``spray'' of material in this match. Such investigation of mass models would make for interesting future work. 

\subsection{The Mice}

\begin{figure*}
\includegraphics[width=\textwidth]{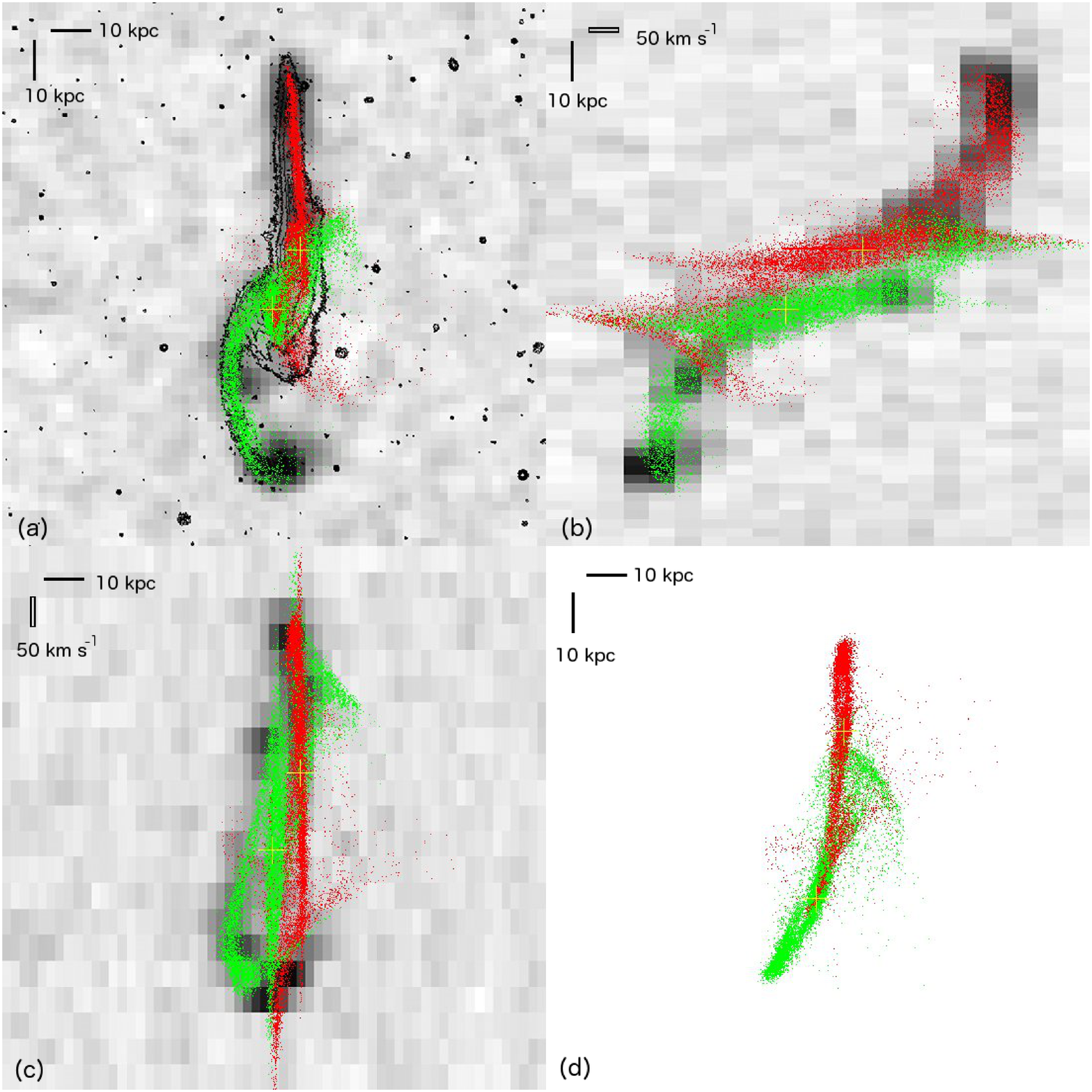}
\caption{Identikit visualization of a self-consistent model for The Mice, matched to the system. (a): sky view of the system ($\alpha$-$\delta$), (b) PV diagram ($v$-$\delta$), (c): PV diagram ($\alpha$-$v$), and (d) ``top-down'' view ($\alpha$,$z$). Here, the sky view is 138 kpc on a side and the velocity width is 903 km s$^{-1}$. In the relevant panels, the solid bar is 10 kpc and the box is $50$ km s$^{-1}$. In the sky view the contours show the R-band image from \citet{Hibbard1996}. Red and green points correspond to stellar particles from N-body realizations representing NGC 4676A and NGC 4676B, respectively. The yellow crosses denote the nuclei of the N-body realizations.}
\label{fig:idkMice}
\end{figure*}

Our best match to this system has two prograde disks, both with small tilts relative to the orbital plane ($i=15^{\circ}$ and $25^{\circ}$ for NGC 4676A and NGC 4676B, respectively; Figure \ref{fig:idkMice}). The initial passage was $r_{peri}=15$ kpc. The two galaxies have just passed for the first time and are still moving towards apocenter. The system is viewed t$_{now}=175$ Myr after first passage and the orbital plane is viewed $\sim20^{\circ}$ degrees from edge on. This, coupled with the inclination of NGC 4647A explains the northern tail's straight appearance. 

The system has baryonic mass $1.3\times10^{11}$ M$_{\odot}$ and dynamical mass of M$_{dyn}\sim6.6\times10^{11}$ M$_{\odot}$. The disks have scale lengths of $3.3$ kpc and circular velocities of $208$ km s$^{-1}$ at 3 disk scale lengths. The similar length of the \HI\ and stellar tails suggests scale lengths for the gas and stellar disks are somewhat similar, though this interpretation for NGC 4676A is complicated by projection effects (the tail is seen edge on).

The observed morphology and kinematics are well matched by our dynamical model, particularly the straightness of the Northern tail and the curvature of the Southern tail. Additionally, the galaxy centers are consistent with both the nuclear positions as measured from 2MASS and the range of systemic velocities obtained from NED. Our model features some bridge material which has penetrated the disks; such material is hinted at in the deep optical images, however these features are not coincident between the model and the data. This may reflect a mass model discrepancy or be an indication that the galaxy inclinations are too low. 

\subsubsection{Comparison with Previous Models}
\label{sssec:MiceModels}

The Mice has several previously published dynamical models. For convenience, some of the parameters from these models have been compiled in Table \ref{table:Micemodels}.

\begin{deluxetable*}{lccccccc}[h!]
\tablecaption{Comparison of dynamical models for the Mice}
\tablehead{\colhead{Source} & \colhead{$e$} & \colhead{$r_{peri}$} (kpc)  & \colhead{NGC 4676A \emph{($i$,$\omega$)}} & \colhead{NGC 4676B \emph{($i$,$\omega$)}} & \colhead{$t$ (Myr)} & \colhead{Self-consistent} & \colhead{3-D Match}}
\startdata
\citet{Toomre1972}	& $0.6$	& \nodata & (15$^{\circ}$, 270$^{\circ}$) & (60$^{\circ}$, 270$^{\circ}$) & $120$ & N & Y\footnote{Velocity was not included in the match, but the relative velocities of the two galaxies was predicted. This prediction was later confirmed by \citet{Stockton1974a}.}\\
\citet{Mihos1993}	& $0.6$	& $23$ & (15$^{\circ}$, 270$^{\circ}$) & (60$^{\circ}$, 270$^{\circ}$) & $180$ & Y & Y \\
\citet{Gilbert1994}	& $1$	& $2.25$ & (20$^{\circ}$, 270$^{\circ}$) & (40$^{\circ}$, 270$^{\circ}$) & $260$ & Y & Y \\
\citet{Barnes2004}	& $1$	& $8.9$ & (25$^{\circ}$, 330$^{\circ}$) & (40$^{\circ}$, 60$^{\circ}$)\phn  & $170$ & Y & Y \\
This paper		& $1$	& $14.8$ & (15$^{\circ}$, 325$^{\circ}$) & (25$^{\circ}$, 200$^{\circ}$) & $175$ & Y & Y
\enddata
\label{table:Micemodels}
\tablecomments{Col 1: Reference for the model. Col 2: Eccentricity of the orbit. Col 3: Pericentric separation in physical units (kpc). Col 4: ($i$,$\omega$) for NGC 4676A. Col 5: ($i$,$\omega$) for NGC 4676B. Col 6: Time since first pericenter passage in physical units (Myr). Col 7: Did the simulation include self-consistently modeled massive dark matter halos and stellar distributions? Col 8: Was the model constrained by three dimensional data ($\alpha$,$\delta$,$v_r$)? Values in physical units have not been adjusted to reflect differences in cosmology.}
\end{deluxetable*}

The initial model of this system by TT72 placed the orbital plane of the $e=0.6$ orbit roughly parallel to our line of sight with a low inclination ($15^{\circ}$) for NGC 4676A and a larger but still prograde inclination ($60^{\circ}$) for NGC 4676B. This configuration was able to reproduce the morphology of the tails as well as the sense of the rotation velocity for the main bodies of the galaxies.

\citet{Gilbert1994} built on the TT72 model by incorporating a self-consistently treated dark matter halo, finding a plausible match to the system only by assuming a $e=1$ orbit. The disk inclinations were similar to those in TT72, but a suitable match required more time elapsed since first pericenter passage. 

The TT72 model was later revised by \citet{Mihos1993} to study the effects of star formation in this system. The same disk inclinations and orbit ($e=0.6$) were used, but match the system at an earlier time. \citet{Mihos1993} noted this has the effect of harming the match to the tidal tails but it improves the match to the main bodies of the galaxies. The time since pericenter is $180$ Myr for this model.

As part of an effort to test numerical star formation prescriptions using The Mice, \citet{Barnes2004} utilized a dynamical model qualitatively similar to previous efforts. The disk inclination for NGC 4676A is slightly higher ($25^{\circ}$) while the inclination for NGC 4676B is somewhat lower ($40^{\circ}$). Utilizing an $e=1$ orbit, the best match to the system comes roughly $170$ Myr after pericenter passage. This model reproduces the length of the tails, but the bar in NGC 4676B is misaligned relative to observations. This misalignment is not be a fatal flaw, particularly if the bar was in place before the encounter began \citep[e.g.,][]{Barnes2004}.

A parameter space survey of the Mice by Barnes (in prep) using the Identikit 2 technique \citep{Barnes2011a} suggests that physical values such as the time since pericenter passage remain is fairly well constrained, with essentially all acceptable matches consistent with a first passage $150-200$ Myr ago, consistent with the model presented here.

In summary, our dynamical model has a similar inclination for NGC 4676A and a somewhat lower inclination for NGC 4676B than has been previously found. The time since pericenter passage (in Myr) is relatively consistent between the various models. Our model reproduces the 3D data for the system, particularly the tidal tails. Tweaks to the mass model may improve discrepancies such as the bar.

\subsection{Antennae}

\begin{figure*}
\includegraphics[width=\textwidth]{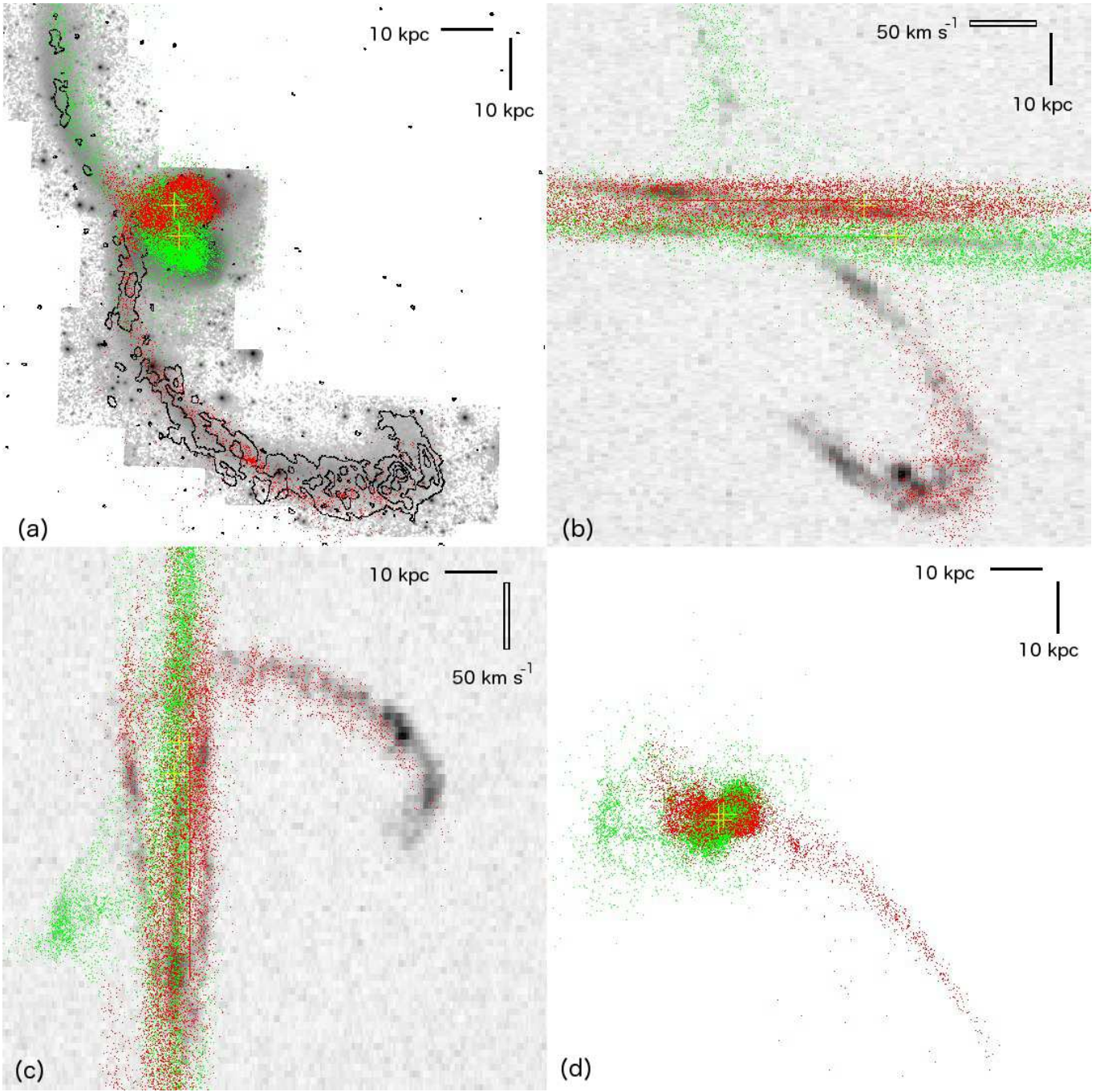}
\caption{Identikit visualization of a self-consistent model for the Antennae, matched to the system. (a): sky view of the system ($\alpha$-$\delta$), (b) PV diagram ($v$-$\delta$), (c): PV diagram ($\alpha$-$v$), and (d) ``top-down'' view ($\alpha$,$z$). The sky view shows \HI\ contours over a V band image, both from \citet{Hibbard2001}; it is 105 kpc on a side. The position velocity diagrams are the maximum pixel of the relevant projection of the cube, with darker pixels corresponding to higher peak emission levels. The velocity width is 410 km s$^{-1}$. In the relevant panels, the solid bar is 10 kpc and the box is $50$ km s$^{-1}$. Red and green dots correspond to stellar particles from N-body realizations of NGC 4038 and NGC 4039, respectively. The yellow crosses denote the nuclear positions of the N-body realizations.}
\label{fig:idkAntennae}
\end{figure*}

Consistent with expectations from visual inspection of the tail morphology and with previous dynamical models for the system, we find an encounter geometry of two prograde disks (Figure \ref{fig:idkAntennae}). The disk angles are $(i,\omega)=(65^{\circ},~345^{\circ})$ and $(70^{\circ},~95^{\circ})$ for NGC 4038 and NGC 4039 respectively. The separation at first pericentric passage is approximately 5 kpc. The system is viewed around its second passage (t$_{now}=260$ Myr after first passage) and the main bodies will merge in approximately 70 Myr.

The system has a total dynamical mass of M$_{dyn}\sim8\times10^{11}$ M$_{\odot}$ of which $1.6\times10^{11}$ M$_{\odot}$ is baryonic. In our model, the disks of both galaxies have scale lengths of $1.5$ kpc and circular velocities of $330$ km s$^{-1}$ at 3 disk scale lengths.

The dynamical model reproduces the general morphology and kinematics of both tidal tails. The hook in the Southern tail is partially reproduced, but not not entirely satisfactorily. The best fit time is around 2nd pericenter passage, however there is some uncertainty regarding the precise timing (i.e., before or after 2nd passage). In these simulations the tidal tails have minor evolution over this relatively short time range, providing limited constraint on the more rapidly evolving central regions.

The modeled velocity extent in the main bodies of the galaxies is quite large (Figure \ref{fig:idkAntennae}), possibly extending beyond the velocity range observed in \HI\ \citep[though this is unclear due to the relatively narrow bandwidth of the observations,][]{Hibbard2001} . If these high velocity components originate as re-accreted tidal material, the gaseous component will likely be shocked upon its return to the disk and we would not see it in \HI.

Finally, in contrast with models for the other three systems presented here, the implied circular velocity of the progenitor disks is significantly higher than typically seen in late type galaxies. It appears that this discrepancy can not be satisfactorily addressed using the mass model assumed here (i.e., there is not another region of parameter space which fits the 3D data and has a more reasonable circular velocity for the progenitors). The Antennae may reflect a system in which the actual mass configuration was sufficiently different from our assumed mass model that it has a major effect on the ability to match a dynamical model. A revised mass model may have more success in producing a match to the data while also providing reasonable progenitor properties.

\subsubsection{Comparison with Previous Models}
\label{sssec:AntennaeModels}

Here we mention the previously published models for the Antennae. For convenience, some of the parameters from these models have been compiled in Table \ref{table:Antennaemodels}.

\begin{deluxetable*}{lccccccc}[h!]
\tablecaption{Comparison of dynamical models for the Antennae}
\tablehead{\colhead{Source} & \colhead{$e$} & \colhead{$r_{peri}$} (kpc)  & \colhead{NGC 4038 \emph{($i$,$\omega$)}} & \colhead{NGC 4039 \emph{($i$,$\omega$)}} & \colhead{$t$ (Myr)} & \colhead{Self-consistent} & \colhead{3-D Match}}
\startdata
\citet{Toomre1972}	& $0.5$	& \nodata & (60$^{\circ}$, 330$^{\circ}$) & (60$^{\circ}$, 330$^{\circ}$) & $150$ & N & N\\
\citet{vanderHulst1979}\footnote{This was a rescaling of the TT72 model, motivated by \HI\ data on the tails.}	& $0.5$	& \nodata & (60$^{\circ}$, 330$^{\circ}$) & (60$^{\circ}$, 330$^{\circ}$) & $50$ & N & N\footnote{Matches the sense of the rotation but not the velocities.}\\
\citet{Barnes1988}	& $1$	& $20$ & (60$^{\circ}$, 330$^{\circ}$) & (60$^{\circ}$, 330$^{\circ}$) & $200$ & N & N\\
\citet{Mihos1993}	& $0.5$	& $27$ & (60$^{\circ}$, 330$^{\circ}$) & (60$^{\circ}$, 330$^{\circ}$) & $210$ & Y & N\\
\citet{Dubinski1996}\footnote{This was a simulation of an Antenna-like system, not specifically the Antennae.}	& $1$ & \nodata & (60$^{\circ}$, 330$^{\circ}$) & (60$^{\circ}$, 330$^{\circ}$) &  \nodata & Y & N\\
Hibbard \& Barnes\footnote{From http://www.cv.nrao.edu/\~jhibbard/n4038/n4038sim/}& $1$ & \nodata & (60$^{\circ}$, 300$^{\circ}$) & (30$^{\circ}$, 120$^{\circ}$) & $220$ & Y & Y\\
\citet{Karl2010}	& $\approx1$	& $10.4$ & (60$^{\circ}$, 30$^{\circ}$)\phn & (60$^{\circ}$, 60$^{\circ}$)\phn & $600$ & Y & Y  \\
This paper		& $1$	& $5$ & (65$^{\circ}$, 345$^{\circ}$) & (70$^{\circ}$, 95$^{\circ}$)\phn & $260$ & Y & Y
\enddata
\label{table:Antennaemodels}
\tablecomments{Col 1: Reference for the model. Col 2: Eccentricity of the orbit. Col 3: Pericentric separation in physical units (kpc). Col 4: ($i$,$\omega$) for NGC 4038. Col 5: ($i$,$\omega$) for NGC 4039. Col 6: Time since first pericenter passage in physical units (Myr). Col 7: Did the simulation include self-consistently modeled massive dark matter halos and stellar distributions? Col 8: Was the model constrained by three dimensional data ($\alpha$,$\delta$,$v_r$)? Values in physical units have not been adjusted to reflect differences in cosmology}
\end{deluxetable*}

The TT72 model utilizes an encounter between two disks with the same disk orientations relative to the orbital plane $(i,\omega)=(60^{\circ},~330^{\circ})$. The orbit is $e=0.5$ and the system is viewed after apocenter. The TT72 model results in equal length tails, in contrast to what is observed. Adjustments to this model by \citet{vanderHulst1979} reduced the inferred masses of the system and the viewing time but maintained the symmetric tails.

\citet{Barnes1988} revisited the Antenna with a simulation that included self-consistently treated dark matter halos. Both Barnes and \citet{Mihos1993} utilized the same parameters as TT72, but employ self-gravitating disks (plus gas in the latter case). The more realistic treatment of the merger results in the much more rapid decay of the orbits when compared with TT72, leading to an earlier viewing time.

\citet{Dubinski1996} modeled the system as part of an exploration of the effect of dark matter halos on the properties of tidal tails. The orbital parameters are similar to \citet{Barnes1988} but the halo:baryonic mass ratio was varied, exploring values of 4:1, 8:1, 15:1, and 30:1. The focus of the exercise was to explore the effect of isothermal dark matter halos on tail formation, so the matches are somewhat poor. In the 4:1 and 8:1 cases sufficiently long tails are formed to be plausible for the Antennae. In the cases of 15:1 and 30:1, the tail length is considerably shorter and the tails even fall back into the progenitors before they merge. \citet{Dubinski1996} conclude the Antennae do not feature such extreme halo masses.

\citet{Karl2010} present a model which differs from previous matches to the kinematics and morphology of the system. This model matches the system at a much later time (after second passage). It has the same disk inclinations as previous matches but different arguments at pericenter. Their model implies more reasonable disk rotation values ($\sim190$ km s$^{-1}$), likely a result of the mass model adopted.

Our model has disk orientations for NGC 4038 which are consistent with early matches to the system, while the disk orientations for NGC 4039 are closer to those from \citet{Karl2010}. The large difference in the time since pericenter compared to our model may be linked to a difference in mass models used; the galaxy mass models used by \citet{Karl2010} are more than twice as dark matter dominated as the models here. This, combined with a wider pericentric separation (as it relates to the disk sizes and the dark matter halo sizes) influences the orbital evolution and thus the duration of the interaction. A ``mixed'' simulation using the mass models from this work and the dynamical model parameters from \citet{Karl2010} was done in order to compare with their results and the results presented here. As compared with this ``mixed'' simulation, both the \citet{Karl2010} model and the model presented here more accurately reproduced the morphology and kinematics of the Antennae, consistent with the differing mass models being the source of the disagreement in dynamical model parameters. 

\subsection{NGC 2623}

\begin{figure*}
\includegraphics[width=\textwidth]{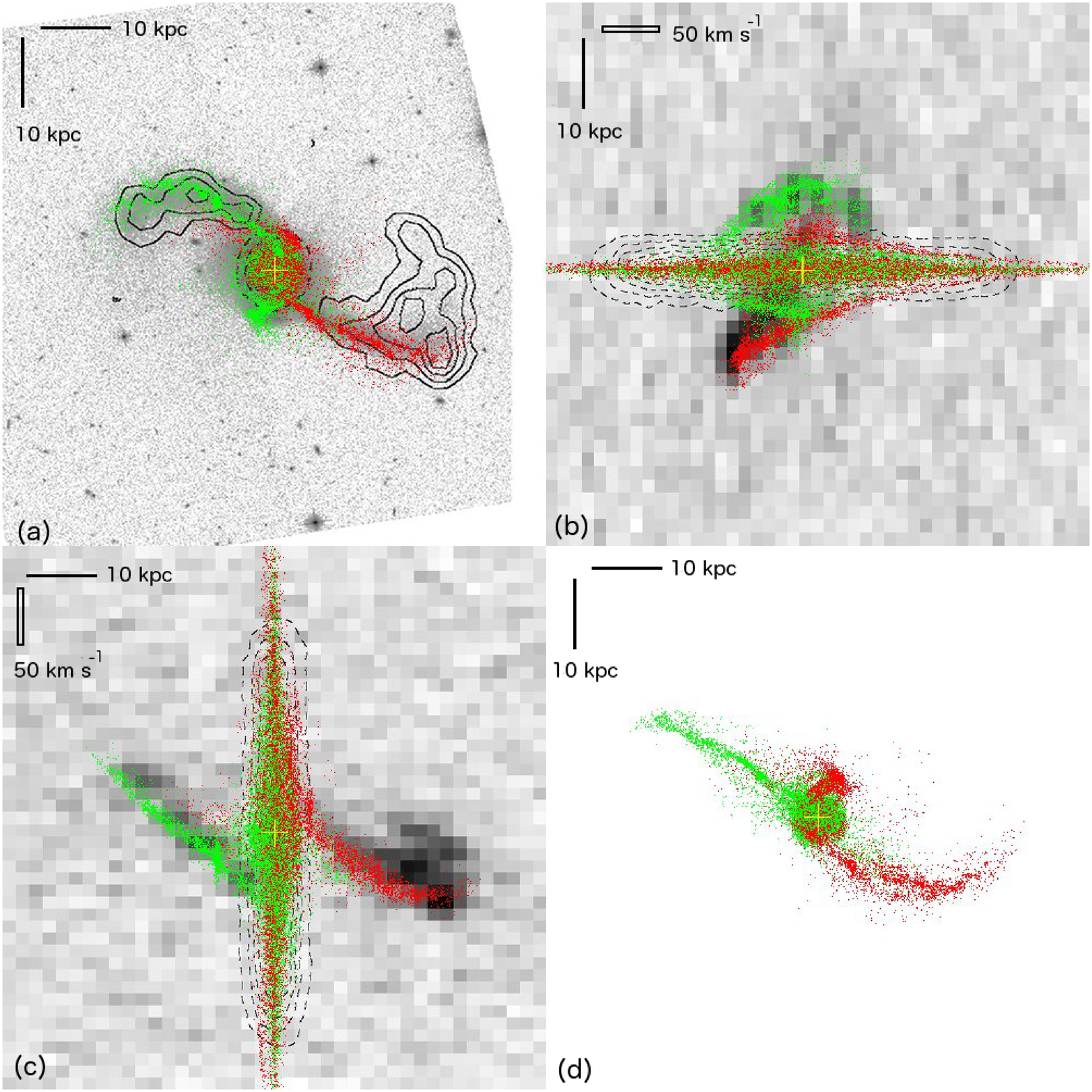}
\caption{Identikit visualization of a self-consistent model for NGC 2623, matched to the system. (a): sky view of the system ($\alpha$-$\delta$), (b) PV diagram ($v$-$\delta$), (c): PV diagram ($\alpha$-$v$), and (d) ``top-down'' view ($\alpha$,$z$). The sky view shows the HST F814W image in grayscale \citep{Evans2008} and the \HI\ in contours; it covers 77 kpc. The position velocity diagrams show the \HI\ emission in grayscale and the \HI\ absorption is shown as dashed contours. The velocity width is 470 km s$^{-1}$. In the relevant panels, the solid bar is 10 kpc and the box is $50$ km s$^{-1}$. Red and green points correspond to stellar particles from N-body realizations of the two progenitor systems for NGC 2623 and the yellow cross denotes the nuclei of the N-body realizations.}
\label{fig:idkN2623}
\end{figure*}

\begin{figure}
\includegraphics[width=0.5\textwidth]{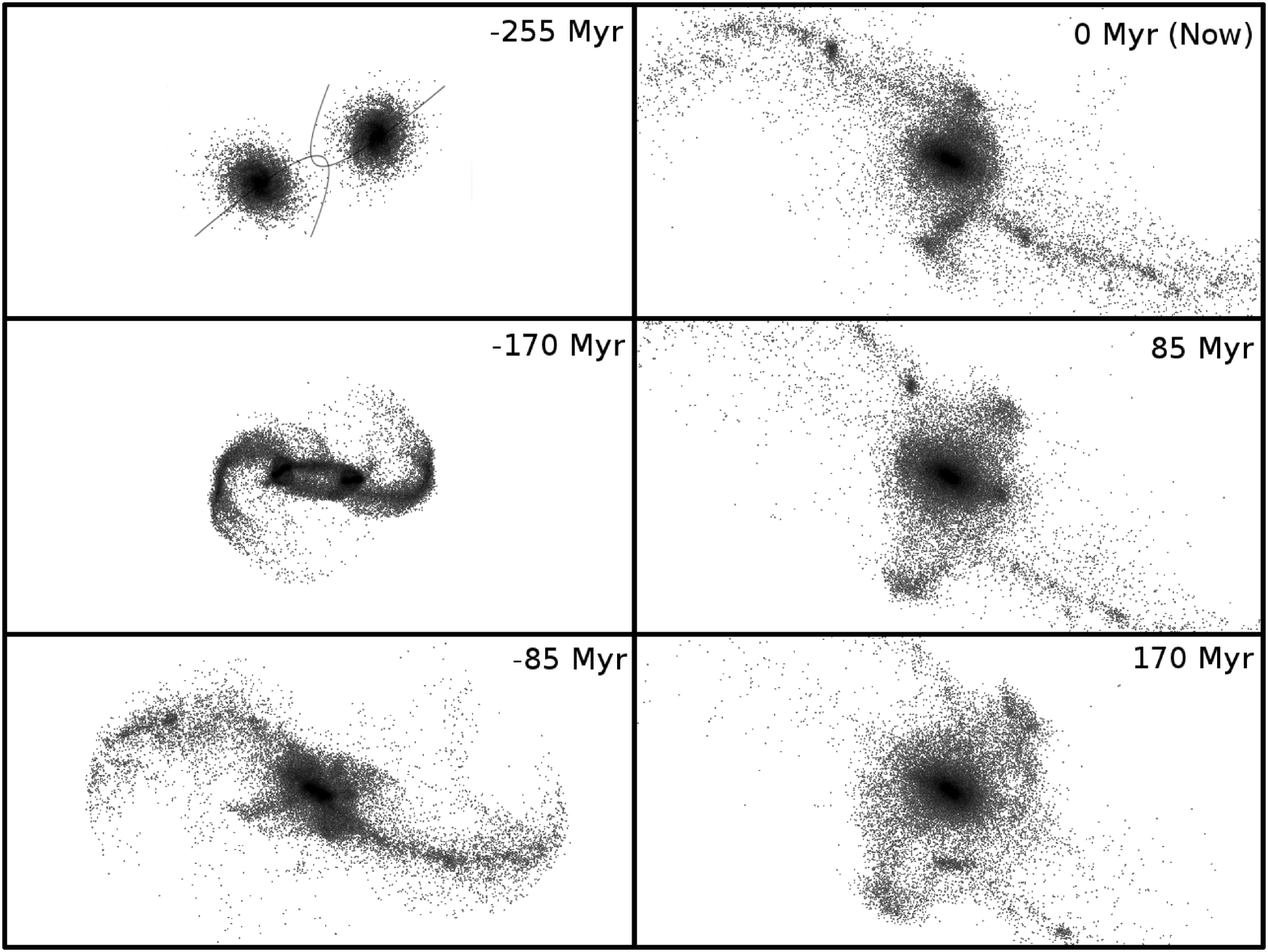}
\caption{Snapshots of the evolution of NGC 2623 as a function of time, based on the dynamical model presented here and shown in the sky plane. The stellar distribution is shown in grayscale in 75 kpc by 50 kpc boxes. The grey lines show the initial Keplerian orbits of the galaxies projected onto the plane of the sky. The times shown are relative to now (cf. Figure \ref{fig:decay}).}
\label{fig:N2623evol}
\end{figure}

NGC 2623 shows prominent tidal tails on either side of a merger remnant. Our simulation is the first dynamical model for this system (Figures \ref{fig:idkN2623} and \ref{fig:N2623evol}). The progenitor disks were both on prograde orbits, with the spin axes have a moderate tilt relative to the orbital plane ($i=30^{\circ}$ and $25^{\circ}$ respectively). The arguments of the disk spins are $\omega=330^{\circ}$ and $110^{\circ}$. The orbital plane itself is close to the plane of the sky and the object is viewed $t_{now}\sim220$ Myr after first pericenter passage. The initial passage was roughly $r_{peri}=0.8$ kpc (Figure \ref{fig:decay}).

Our matches require relatively compact progenitor systems, with disk scale lengths less than $1$ kpc, circular velocities of $153$ km s$^{-1}$ at three scale lengths, and a dynamical mass of M$_{dyn}\sim6\times10^{10}$ M$_{\odot}$ for the system. Our mass model then implies a baryonic mass of $\sim10^{10}$ M$_{\odot}$. Stellar mass estimates from \citet{U2012} range from $0.9-2.3\times10^{10}$  M$_{\odot}$, consistent with our estimate. We experimented with merger simulations of a 2:1 mass ratio, but generally found better agreement with the morphology and kinematics using an equal-mass encounter.

This dynamical model has the system being viewed after the nuclei have coalesced. The tidal tails are well reproduced, though the full extent of the SW tail is not reproduced. In the model, this tail does begin to curl around; it is possible that a more extended gas disk would reproduce the more extended \HI\ emission in the tail. The velocity width of the model in the merged body is significantly wider than the \HI\ absorption which may be due to a comparison of non-dissipational particles with a diagnostic which is inherently dissipational.

This system is being viewed after some of the tidal tail material has turned around and begun falling back into the merger remnant, as expected. In particular, a loop of material has formed south of the nucleus (Figure \ref{fig:N2623evol}) from material returning from the NE tidal tail.

Many of the optically visible clusters in NGC 2623 are located just south of the nucleus. \citet{Evans2008} suggested the activity in this region was related to the return of material from the tidal tails. This prediction is consistent with the result of our simulations which show that at this point in the merger stage, the more tightly bound material in the tails (i.e., the material near the base of the tails) has already begun falling back into the remnant. While a detailed comparison requires the inclusion of gas, at this point it seems plausible that the infall of tidal material through this region is the trigger of the off-nuclear star formation; as the tails fall in, they loop around the nucleus and the off-nuclear clusters coincide with a turnaround point for the Eastern tail on its path towards the nucleus. The nuclear star formation is likely fueled by a combination of material returning from the tidal tails and gas still present in the main bodies of the galaxies during their final merger, though verification will require tests from simulations incorporating gas dynamics.

\subsection{Future Work}
\label{sec:Discussion}

The four dynamical models presented here provide both a validation of the Identikit method as applied to real data and establish the placement of these objects on a merger timeline (Figure \ref{fig:decay} and Table \ref{table:models}). 

Dynamical models are useful for studies of individual systems. Coupling them with a prescription (or accurate physical description) for star formation can allow one to predict the evolution of a system in different tracers of activity. Doing this for a statistically significant sample of systems can establish whether gas-rich mergers in the local Universe all undergo an Ultra Luminous Infrared Galaxy (ULIRG) phase or if such luminous objects require special progenitor properties or encounter geometries. Prescriptions of AGN feedback can potentially be tested as well, to examine the analogous question for quasar activity. The ease of matching galaxy mergers with the Identikit technique should enable the construction of samples of tens of systems with dynamical models to facilitate examination of these questions.

By design the self-consistent simulations presented here do not include a dissipational component. Observationally, all of these systems show elevated levels of star formation; a natural next step is to run N-body+SPH simulations to include the effects of gas. It would be instructive to see if these simulations can replicate the morphology and kinematics as determined from \HI\ and CO observations.

A study including gas should also incorporate star formation. Star formation prescriptions in simulations are uncertain and the details can have an effect on those cosmological simulations which explicitly include gas. An accurate prescription would reproduce the observed properties of star formation in these systems. A number of options are available to compare, including global tests (e.g., the global star formation rate) and resolved tests (e.g., spatial distribution of ongoing star formation via comparison with H$\alpha$ or $8~\mu$m maps, or properties of observed star clusters). 

To date, most numerical studies of star formation in mergers utilize a prescription tuned to match the Schmidt law \citep[SFR $\propto\rho^{n}$,][]{Schmidt1959}. Studies of two systems have suggested that the addition of a shock criteria to star formation prescriptions more accurately reproduces observed properties of young star clusters in merging systems \citep[for The Mice and NGC 7252, respectively]{Barnes2004,Chien2010}. A detailed numerical treatment of star formation by \citet{Hopkins2011d,Hopkins2012} incorporates various feedback mechanisms in high-resolution simulations to recover the observed star formation efficiency of the Schmidt-Kennicutt Law, without tunable parameters. 

The duty cycles of LIRGs and ULIRGs remains an open question --- i.e., will every massive, gas-rich galaxy merger undergo a ULIRG phase? If not, what properties (orbital or galaxy structure) determine whether a merger will undergo a ULIRG phase? Answering this question is limited by the numerical treatment of star formation.

Comparing the observed star formation properties with those predicted from star formation prescriptions in detailed simulations matched to the systems may constrain these prescriptions and provide guidance for cosmological simulations. The dynamical models present here provide a useful starting point for exploring these questions.

\section{Summary}
\label{sec:summary}

Using Identikit's hybrid simulations and visualization tool, we have obtained dynamical models for NGC 5257/8, The Mice, the Antennae, and NGC 2623 which reproduce the observed morphology and kinematics of the tidal features and place them on a merger timeline. Follow-up simulations with self-gravitating disks verify the matches and generally show agreement with self-gravitating features such as spiral arms. The dynamical models are all consistent with these systems being roughly equal mass encounters of disk galaxies involved in prograde interactions. 

The models presented here are the first to appear in the literature for NGC 5257/58 and NGC 2623, and the new models for the Mice and the Antennae are compared with previously published models. Based on the assumed mass model and initial conditions, the results indicate the four systems are currently being viewed 175-260 Myr after first pericenter passage, and they have estimated times to coalescence of 1200 Myr (NGC 5257/58), 775 Myr (Mice), 70 Myr (Antennae) and -80 Myr (NGC 2623). Future simulations incorporating gas will be used to test numerical prescriptions of star formation.

\acknowledgements

The authors thank Francois Schweizer for helpful comments on the manuscript. The authors also thank the anonymous referee for useful comments. G.C.P. and A.S.E. were supported by NSF grants AST 1109475 and 02-06262, and by NASA through grants HST-GO10592.01-A and HST-GO11196.01-A from the Space Telescope Science Institute, which is operated by the Association of Universities for Research in Astronomy, Inc., under NASA contract NAS5-26555. This research has made use of the NASA/IPAC Extragalactic Database (NED) which is operated by the Jet Propulsion Laboratory, California Institute of Technology, under contract with the National Aeronautics and Space Administration. This research has made use of NASA's Astrophysics Data System. The National Radio Astronomy Observatory is a facility of the National Science Foundation operated under cooperative agreement by Associated Universities, Inc.



\end{document}